\documentclass[sigconf]{acmart}
\AtBeginDocument{%
  }

\usepackage{amsmath,amsfonts}
\usepackage{graphicx}
\usepackage{float}
\usepackage[ruled,vlined,linesnumbered]{algorithm2e}
\usepackage{algpseudocode}
\usepackage{graphicx}
\usepackage{textcomp}
\usepackage{cleveref}
\usepackage{url}
\usepackage{xspace}
\usepackage{multirow}
\usepackage{booktabs}
\usepackage{enumitem}
\usepackage{pifont}

\newcommand{\tool}{\texttt{LagEase}\xspace}

\author{Rui Lu}
\affiliation{%
    \institution{East China Normal University}
    \city{Shanghai}
    \country{Shanghai}
}
\email{ruilu@stu.ecnu.edu.cn}
\setcopyright{rightsretained}
\copyrightyear{2025}
\acmYear{2025}
\acmDOI{XXXXXXX.XXXXXXX}
\acmConference[ICSE-Companion '25]{The 47th International Conference on Software Engineering}{April 27-May 3, 2025}{Ottawa, Canada}
\acmISBN{978-1-4503-XXXX-X/2018/06}

\title{Towards Compatibly Mitigating Technical Lag in Maven Projects}

\begin{document}

\begin{abstract}
Library reuse is a widely adopted practice in software development, 
however, 
re-used libraries are not always up-to-date, thus including unnecessary bugs or vulnerabilities. 
Brutely upgrading libraries to the latest versions is not feasible because breaking changes and bloated dependencies could be introduced, which may break the software project or introduce maintenance efforts. Therefore, balancing the technical lag reduction and the prevention of newly introduced issues are critical for dependency management.
To this end, \tool is introduced as a novel tool designed to address the challenges of mitigating the technical lags and avoid incompatibility risks and bloated dependencies.
Experimental results show that \tool outperforms Dependabot , providing a more effective solution for managing Maven dependencies.
\end{abstract}

\maketitle

\section{Introduction}
The reuse of Third Party Libraries (TPLs) is common in software development. 
However, the TPLs cannot always be guaranteed to be the latest versions due to several concerns, such as stability, which may lead to unfixed bugs, potential vulnerabilities, and under-utilized features. 
To represent the degree of obsolescence of a library, González-Barahona et al.~\cite{gonzalez2017technical} proposed the concept of technical lag, which indicates the latency between the latest available snapshot and the current artifact in use. Zerouali et al.~\cite{zerouali2019formal} proposed a formula for calculating technical lag as the interval of time or versions and unveiled the prevalence of technical lags. 
Though exposed, unfortunately, the technical lags have not been systematically resolved by existing tools to keep dependencies up-to-date. Upgrading dependencies, however, is not straightforward, which could introduce breaking changes, leading to failed compilation, runtime error, and unexpected output\cite{bavota2013evolution,kong2024towards}. Additionally, upgrading dependencies could significantly modify the dependency graph structures, involving bloated and redundant dependencies\cite{soto2021comprehensive,song2024efficiently,soto2021longitudinal}.
The above-mentioned issues should be avoided when mitigating the technical lag.

To address this gap, I have developed a novel tool \tool that systematically considers all dependencies in a proper order to mitigate the overall technical lags. During the mitigation, the compatibility and dependency debloating are considered for smooth upgrading by implementing code-centric program analysis.

\section{Related Work}
Revealed by Stringer et al.~\cite{stringer2020technical} , 
The majority of fixed versions are outdated, increasing technical lag substantially. 
Cox et al.~\cite{cox2015measuring} found that projects using outdated dependencies were four times more likely to have security issues than those with up-to-date dependencies. 
Additionally, dependencies included transitively are more likely to introduce vulnerabilities\cite{lauinger2018thou}. 
These studies inspired us to upgrade all dependencies, including transitive ones to reduce technical lag.

\begin{figure*}[ht]
\centering
\includegraphics[width=0.6\linewidth]{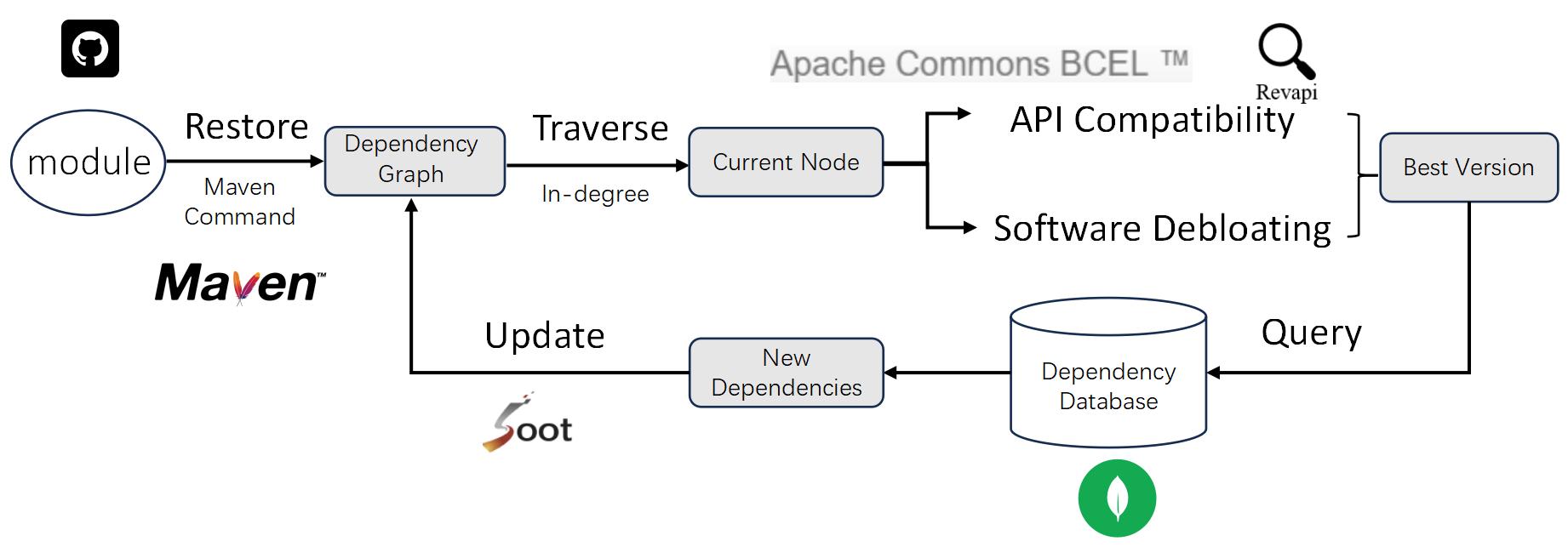}
\caption{Overview of \tool}
\label{fig:tool}
\end{figure*}

\begin{table*}[ht]
\caption{Evaluation Results}\label{tab:result}
\setlength{\tabcolsep}{4pt}
\resizebox{\textwidth}{!}{
\begin{tabular}{@{}lllllllll@{}}
\toprule
        &   \textbf{\#Module} &  \textbf{Compile Failure} & \textbf{Test Failure} & \textbf{Intact Module}&\textbf{Original Tech lag} &\textbf{Reduced Tech Lag} & \textbf{Original Dep} & \textbf{Reduced Dep} \\\midrule
\textbf{ \tool} & 182 & 0 & 5  & 177 & 9,866 & 7,887 & 1,232  & 5\\   
\textbf{Dependabot} & 182 & 0 & 6 & 176 & 9,866 & 665 & 1,232 & 0\\ \bottomrule
\end{tabular}
}
\end{table*}

\section{Methodology}
\tool is designed for Java projects from Maven~\cite{Maven}. 
Generally, \tool first restores the original dependency graph from a given project and then traverses the graph to mitigate its technical lag while maintaining the compatibility and dependencies unbloated. After each iteration, the dependency graph is dynamically updated. 

\subsection{Restoring Original Dependency Graph}
For each Java project, 
given the initial tree returned by Maven command~\cite{MavenTree}, \tool restores the hidden edges among valid nodes (\texttt{Test} and \texttt{Provided} dependencies are excluded~
\cite{zhang2023compatible}) to derive a complete dependency graph.

\subsection{Traversing Dependency Graph}

\tool follows a systematic order to traverse the dependency graph, similar to topological sorting.\tool first filters the candidate versions using two constraints to identify those that disobey dependency debloating and compatibility requirements.
For dependency debloating, \tool iteratively counts dependencies of all versions in the candidate versions using the Maven\cite{Maven}, then, removes versions with more dependencies than the original version from candidates.
For compatibility,
because incompatibility depends on the context of specific user projects, relying on breaking APIs of TPLs is not accurate in predicting compatibility. \tool implements a usage analysis of the breaking APIs detected by Revapi~\cite{revapi} by conducting reachability and reference analysis of Java constructs~\cite{JSL}, such as classes, methods, and fields. The reachable and used constructs are calculated with Soot~\cite{Soot} and BCEL~\cite{BCEL} based on Java bytecode by tracing the def-use and call chains. If detected breaking APIs and constructs overlap with the used ones, the user project is considered affected by incompatibility risks, thus excluding from candidates.

After the filtering, \tool selects the latest version from the remaining candidate versions as the optimal version for this node. 
\subsection{Update Graph}

After upgrading a node, \tool updates the node in the dependency graph with other relevant downstream nodes to ensure that the next node relies on the up-to-date context.

\section{Evaluation and results}

182 successfully compiled and tested modules were gathered from 5 repositories(out of a total of 270 modules) as the dataset.
Due to the lack of similar tools, Dependabot\cite{Dependabot} as the GitHub dependency management tool that automatically upgrades dependencies served as a baseline, although its target is not about technical lags. The results of \tool and Dependabot are shown in \Cref{tab:result}.
\tool breaks fewer modules and reduces more technical lag and dependencies than Dependent does.
\section{Conclusion and contributions}
I proposed \tool, a tool designed to upgrade dependencies in Maven projects to reduce technical lag, while maintaining compatibility and software debloated. A comparison with Dependabot reveals that \tool outperformed Dependabot at multiple metrics. Although \tool is designed for Maven projects, its principles can also be applied to other ecosystems, such as npm.

\bibliographystyle{ACM-Reference-Format}
\bibliography{reference}
\end{document}